\documentclass[%
reprint,
superscriptaddress,
amsmath,
amssymb,
aps,
pra,
]{revtex4-1}
\usepackage{amssymb,amsmath}
\usepackage{mathrsfs}
\usepackage{braket}
\usepackage{color}
\usepackage{graphicx}
\usepackage{dcolumn}
\usepackage{bm}
\usepackage{float}

\usepackage{tikz}

\usepackage{ulem}

\usepackage[hypertexnames=false,colorlinks=true,linkcolor=red,citecolor=blue,urlcolor=blue]{hyperref} 
\usepackage{natbib} 

\begin{document}

\title{Magnon cat states in a cavity-magnon-qubit system via two-magnon driving \\ and dissipation}

\author{Gang Liu}
\email{gangliu@zju.edu.cn}
\affiliation{Zhejiang Key Laboratory of Micro-Nano Quantum Chips and Quantum Control, School of Physics, and State Key Laboratory for Extreme Photonics and Instrumentation, Zhejiang University, Hangzhou 310027, China}

\author{Gen Li}
\affiliation{Zhejiang Key Laboratory of Micro-Nano Quantum Chips and Quantum Control, School of Physics, and State Key Laboratory for Extreme Photonics and Instrumentation, Zhejiang University, Hangzhou 310027, China}

\author{Huatang Tan}
\email{tht@mail.ccnu.edu.cn}
\affiliation{Department of Physics, Huazhong Normal University, Wuhan 430079, People's Republic of China}

\author{Jie Li}
\affiliation{Zhejiang Key Laboratory of Micro-Nano Quantum Chips and Quantum Control, School of Physics, and State Key Laboratory for Extreme Photonics and Instrumentation, Zhejiang University, Hangzhou 310027, China}


\begin{abstract}
	
	We propose an efficient method for dissipative generation of magnonic cat states in a cavity-magnon-qubit hybrid system by exploiting a two-magnon driving and dissipation mechanism. When both the magnon and qubit are driven, a coherent nonlinear two-magnon interaction is induced, wherein the qubit and the magnon mode exchange energy through magnon pairs. The dissipation of the qubit is exploited to steer the magnon mode into a quantum superposition of distinct coherent states, where the magnon mode evolves into either an even or odd cat state, depending on the parity of the magnon initial state. For the case where the magnon initial state is a superposition state, e.g., of $\ket{0}$ and $\ket{1}$, the magnon mode can evolve into a weighted mixture of the even and odd cat states. We also find that magnon squeezed states may emerge during the short-time evolution, showcasing the capability of our mechanism in preparing diverse magnon non-classical states.   Magnonic cat and squeezed states are macroscopic quantum states and find applications in macroscopic quantum studies and quantum sensing, e.g., in the dark matter search using ferromagnetic axion haloscopes.
	
\end{abstract}  


\maketitle


\section{\label{sec:level1}INTRODUCTION}

Hybrid quantum systems based on magnonics have received considerable attention in the past decade due to their potential important applications in quantum information science, quantum technology, quantum sensing, nonlinear and macroscopic quantum studies \cite{lachance2019hybrid,yuan2022quantum,Rameshti-2022,zuo2024cavity}. 
Magnons are collective spin excitations in magnetic materials, e.g. yttrium iron garnet (YIG). The magnonic systems based on YIG have many excellent properties, such as high spin density, low dissipation rate, great tunability, etc. 
In particular, one distinct advantage of the magnonic system is that it can coherently couple with diverse quantum systems, including microwave photons, optical photons and vibration phonons through the magnetic-dipole \cite{Huebl2013prl,Tabuchi2014prl,Zhang2014prl}, magneto-optical \cite{hisatomi2016bidirectional,osada2016cavity,zhang2016optomagnonic,Haigh2016prl}, and magnetostriction interaction \cite{zhang2016cavity,li2018magnon,Potts2021prx,shen2022prl}, respectively. 
Moreover, it can also couple to a superconducting qubit. Experimental efforts have demonstrated indirect coupling between magnons and superconducting qubits via virtual photons of the microwave cavity \cite{tabuchi2015coherent,lachance2017resolving,lachance2020entanglement,wolski2020dissipation,xu2023quantum}, while theoretical proposals indicate that a direct coupling is also possible \cite{Kounalakis2022prl}. Integrating magnons with superconducting qubits forms a compelling hybrid system, which finds broad applications in quantum information science and technology \cite{lachance2019hybrid,yuan2022quantum}, and quantum sensing \cite{wolski2020dissipation}, and for preparing magnonic quantum states via quantum control of the superconducting qubit \cite{lachance2020entanglement,xu2023quantum}. 
In such a cavity-magnon-qubit system, significant experimental progress has been made including, e.g., the realization of the magnon-qubit strong coupling \cite{tabuchi2015coherent}, single-shot detection of a single magnon \cite{lachance2020entanglement}, the superposition state of a single magnon and vacuum \cite{xu2023quantum}, etc. These successful experimental demonstrations have stimulated quantum studies in such a hybrid system.
To date, many theoretical proposals have been offered to prepare magnonic quantum states in the system, including magnon blockade \cite{liu2019prbblockade,xie2020prablockade,wu2021phase,Jin2023prablockade}, entanglement \cite{qi2022generation,ren2022long}, squeezing \cite{2023guoprasqueezed}, cat states \cite{Kounalakis2022prl,he2023pramc,hou2024robust,he2024pracats}, and so on.
The aforementioned studies indicate that the magnon-qubit system is a promising new platform for preparing magnonic quantum states on the macroscopic scale.

Here, we offer a new approach in the cavity-magnon-qubit system for preparing magnonic cat states by exploiting a two-magnon driving-dissipation mechanism. We note that, so far, several other approaches have been provided for preparing magnonic cat states, e.g., by utilizing magnetic anisotropy in ferromagnetic insulators \cite{Sharma2021prbcat}, directly coupling to a transmon qubit via magnetic flux \cite{Kounalakis2022prl,he2024pracats}, performing non-Gaussian operations and exploiting optomagnonic entanglement \cite{sun2021prl}, applying a Floquet drive onto the magnon mode \cite{he2023pramc}, introducing photon parametric coupling \cite{liu2024magnoncat}, and coupling to a superconducting flux qubit via the persistent current \cite{hou2024robust}. 
In our protocol, the approach relies on inducing both transverse and longitudinal couplings via applying two drives to the qubit and an additional drive to the magnon mode.
The interplay between these two couplings results in a strong two-magnon nonlinear interaction, allowing the magnon mode to exchange energy with the qubit by means of magnon pairs.
We show that the spontaneous emission of the qubit can be utilized to steer magnetic excitations into a quantum superposition of distinct coherent states, i.e., a cat state, where the cat state can be either odd or even depending on the parity of the magnon initial state. In addition, we find that when the initial state is a superposition of the vacuum $|0\rangle$ and the single-magnon state $|1\rangle$, with a relative phase of $ \pi/2$, the interference fringes vanish, resulting in a cat state being a mixture of two coherent states losing its coherence.
It is also worth noting that a dynamical squeezed state may emerge during the short-time evolution. 


The paper is organized as follows: In Sec.~\ref{sec:2}, we introduce the system and the Hamiltonian describing the transverse and longitudinal couplings between the qubit and the magnon mode. In Sec.~\ref{sec:3}, we apply an additional qubit drive and derive an effective Hamiltonian that leads to a strong two-magnon nonlinear interaction. We further show that this two-magnon interaction can lead the magnon mode to a steady-state cat state. Finally, we draw the conclusions in Sec.~\ref{sec:4}.

\begin{figure}[t]
	\centering
	\includegraphics[width=0.9\linewidth]{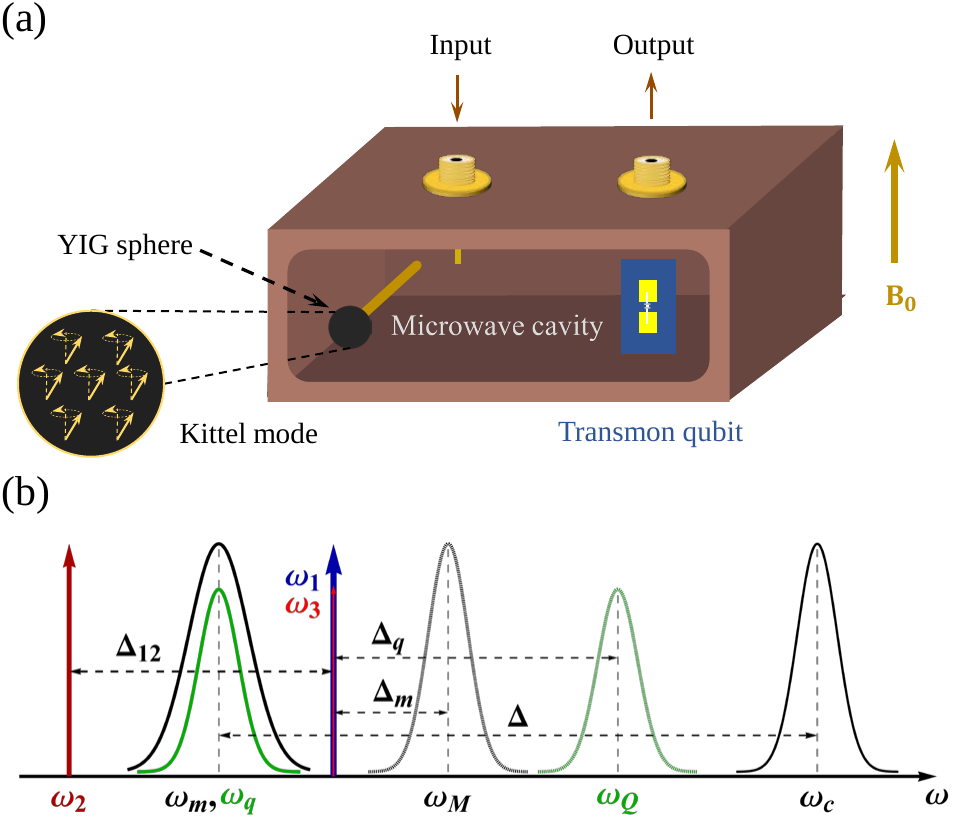}
	\caption{(a) Schematic of the cavity-magnon-qubit system, which consists of a microwave cavity coupled to both a magnon mode of a macroscopic YIG sphere and a superconducting qubit. The YIG sphere, placed in a uniform bias magnetic field along the $z$-axis, is driven by a microwave field. In the meantime, the superconducting qubit is driven by three microwave fields. The magnon mode and the qubit get effectively coupled via the mediation of the microwave cavity.
		(b) Mode and drive frequencies of the system. The cavity with frequency $\omega_c$ is far detuned from the magnon mode with frequency $\omega_m$ and the qubit  with frequency $\omega_q$. Here, $\omega_M$ ($\omega_Q$) is the effective frequency of the magnon mode (the qubit) by eliminating the cavity. See text for details of the multiple driving fields.}
	\label{fig:Fig_1}
\end{figure}

\section{Model and Hamiltonian\label{sec:2}}

The hybrid cavity-magnon-superconducting-qubit system under study is depicted in Fig.~\ref{fig:Fig_1} (a), which consists of a single-crystal YIG sphere and a transmon-type superconducting qubit that are placed inside a microwave cavity. A uniform magnetic field is applied along the $ z $-axis to magnetize the YIG sphere to saturation. The system involves a magnon mode, e.g., the Kittle mode, of the YIG sphere with frequency $ \omega_m $, a superconducting qubit with frequency $ \omega_q $, and a microwave cavity mode with frequency $ \omega_c $. The Kittel mode couples to the microwave cavity via the magnetic-dipole interaction and the latter further couples to the qubit via the electric-dipole interaction. The Hamiltonian of this system reads (setting $\hbar = 1$)  
\begin{align}\label{eq:fullH}
	H =&\ \omega_c c^\dag c + \frac{\omega_q}{2}\sigma_z + \omega_m m^\dag m + g_q(c\sigma_+ + c^\dag \sigma_-)\\
	 +&\ g_m(c m^\dag + c^\dag m),\nonumber
\end{align}
where $c\ (c^\dag)$ and $m\ (m^\dag)$ are the annihilation (creation) operators of the microwave cavity and the magnon mode, respectively, and $\omega_c$ and $\omega_m$ are their resonance frequencies, $\sigma_z = \ket{e}\bra{e} - \ket{g}\bra{g}$ with the excited state $\ket{e}$ and the ground state $\ket{g}$ of the qubit, $\sigma_- = \ket{g}\bra{e}$ and $\sigma_+ = \sigma_-^\dag$ are the lowering and raising operators of the qubit. The coupling rates $g_q$ and $g_m$ are of the cavity-qubit and cavity-magnon systems, respectively.

When the magnon mode and the qubit are (nearly) resonant, $\omega_q \simeq \omega_m \equiv \omega$ , and when they are far detuned from the cavity mode, i.e., $\Delta = \omega_c - \omega \gg g_{m(q)}$, coherent exchange of the qubit excitation and a magnon mode is mediated by virtual photons of the microwave cavity. This allows us to adiabatically eliminate the cavity mode and obtain the effective Jaynes-Cummings-type Hamiltonian of the magnon-qubit system \cite{lachance2019hybrid,2023guoprasqueezed}, which is given by
\begin{align}
	H_{jc} = \omega_M m^\dag m + \frac{\omega_Q}{2}\sigma_z + G (\sigma_+ m + \sigma_- m^\dag),
\end{align}
where $\omega_Q = \omega + g_q^2/\Delta$ and $\omega_M = \omega + g_m^2/\Delta$ correspond to the effective frequencies of the qubit and the magnon mode, respectively [cf. Fig.~\ref{fig:Fig_1}(b)], and $G = g_q g_m/\Delta$ denotes the effective magnon-qubit coupling strength. 
We consider experimentally feasible parameters \cite{tabuchi2015coherent, tabuchi2016quantum, lachance2017resolving}: $\omega/2\pi = 8.1659$~GHz, $\omega_c/2\pi = 8.420$~GHz, $g_m/2\pi = 21$~MHz, and $g_q/2\pi = 121$~MHz. The effective frequencies are then $\omega_Q/2\pi = 8.22352$~GHz and $\omega_M/2\pi = 8.16764$~GHz, with the effective coupling $G/2\pi = 10$~MHz corresponding to the detuning $\Delta =2\pi\times 254.1\ \text{MHz}$. 


To generate a magnon cat state, we apply two transverse microwave Rabi drive tones (with frequencies $\omega_1$ and $\omega_2$) to the qubit and an additional drive (with frequenciy $\omega_3$) to the magnon mode \cite{ballester2012quantum,braumuller2017analog}. In the rotating frame of the dominant drive frequency $\omega_1$, the system yields an effective quantum Rabi model with an added bias term  and a residual effective drive. The effective mode energies in the engineered Hamiltonian are determined by the parameters of the Rabi drives. In the laboratory frame, with all drives applied, the system is described by a Jaynes-Cummings Hamiltonian in the form
\begin{align}\label{eq:full}
	H_1 &=\omega_M m^\dag m + \frac{\omega_Q}{2}\sigma_z 
	+ G (\sigma_+ m + \sigma_- m^\dag)\\
	&+ {\tilde{\eta}_1} \cos(\omega_1 t + \phi_1)\sigma_x
	 + {\tilde{\eta}_2} \cos(\omega_2 t + \phi_2)\sigma_x\nonumber\\
	&+ \eta_3  \cos(\omega_3 t + \phi_3)(m^\dagger + m)\nonumber,
\end{align}
{where $\tilde{\eta}_1 = \eta_1 + \varepsilon_1 \cos(\omega_p t)$ and $\tilde{\eta}_2 = \eta_2 + \varepsilon_2 \cos(\omega_p t)$ are the amplitude-modulated transverse drive strengths applied to the qubit, and $\eta_3$ is the constant drive amplitude applied to the magnon mode, and $\omega_i$ denote the frequencies of the respective driving fields.} In the following,
we set $\phi_i=0$ without loss of generality. 
%
Moving into the rotating frame of the 
drive frequency $\omega_1$, we perform a unitary transformation $U_1 = \exp[i\omega_1(m^\dag m + \sigma_z/2)]$, which leads to the following transformed Hamiltonian:
\begin{align}\label{eq:h1}
	H_2 &=U_1 H_1 U_1^\dag - i U_1 \dot{U}_1^\dag\\
	&= \Delta_m m^\dag m + \frac{\Delta_q }{2}\sigma_z 
	+ G (\sigma_+ m + \sigma_- m^\dag)\nonumber\\
	&+ \frac{{\tilde{\eta}_1}}{2} \sigma_x + \frac{{\tilde{\eta}_2}}{2}(\sigma_+  e^{i\Delta_{12} t} + \sigma_-  e^{-i\Delta_{12} t})\nonumber\\
	& + \frac{\eta_3}{2} (m^\dagger e^{i\Delta_{13} t} + m e^{-i\Delta_{13} t})\nonumber\\
	& + \frac{{\tilde{\eta}_1}}{2} (\sigma_+ e^{2i\omega_1 t}  + \sigma_- e^{-2i\omega_1 t})\nonumber\\
	& + \frac{{\tilde{\eta}_2}}{2} (\sigma_+  e^{i(\omega_1 + \omega_2) t} + \sigma_-  e^{-i(\omega_1 + \omega_2) t})\nonumber\\
	& + \frac{\eta_3}{2} (m^\dagger e^{i(\omega_1 + \omega_3) t} + m e^{-i(\omega_1 + \omega_3) t})\nonumber,  
\end{align}
where $\Delta_m = \omega_M - \omega_1$, $\Delta_q = \omega_Q - \omega_1$, $\Delta_{12} = \omega_1 - \omega_2$, and $\Delta_{13} = \omega_1 - \omega_3$. Applying the rotating wave approximation and setting  $\omega_1 = \omega_3$, we discard the fast-oscillating terms when $\eta_1 / 2\omega_1 \ll 1$. This leads to the following simplified Hamiltonian:
\begin{align}
	H_2^\prime &= H_0 + \Delta_m m^\dag m + \frac{\Delta_q }{2}\sigma_z 
	+ G (\sigma_+ m + \sigma_- m^\dag)\\
	& + \frac{{\tilde{\eta}_2}}{2}(\sigma_+  e^{i\Delta_{12} t} + \sigma_-  e^{-i\Delta_{12} t}) + \frac{\eta_3}{2} (m^\dagger + m )\nonumber\\
	&\, { + \, \frac{\varepsilon_1}{2} \cos(\omega_p t)\sigma_x}\nonumber,
\end{align}
where $H_0 = \eta_1 \sigma_x/2$ represents the dominant static drive component, which transforms the original driving term into a time-independent form.  We now move the remaining terms perturbatively in the interaction picture with respect to $ H_0 $, which results in $ H_3 = e^{i H_0 t} (H_2^\prime - H_0) e^{-i H_0 t} $, in the spin basis $\ket{\pm} = (\ket{g} \pm \ket{e}) / \sqrt{2}$, given by
\begin{align}\label{eq:H3}
	H_3 
	&=\Delta_m m^\dagger m + \frac{\Delta_q}{2} \left(e^{i\eta_1 t} \ket{+}\bra{-} + \text{H.c.}\right) \\
	&+ \frac{G}{2} \left[\left(\ket{+}\bra{+} - \ket{-}\bra{-} - e^{i\eta_1 t}  \ket{+}\bra{-}\right.\right.\nonumber\\
	&\left.\left.\quad + e^{-i\eta_1 t} \ket{-}\bra{+}\right)  m + \text{H.c.}\right]\nonumber \\
	&+ \frac{{\tilde{\eta}_2}}{4}\left[\left(\ket{+}\bra{+} - \ket{-}\bra{-} - e^{i\eta_1 t} \ket{+}\bra{-}\right.\right.\nonumber\\
	&\left.\left.\quad  + e^{-i\eta_1 t}\ket{-}\bra{+}\right) e^{i\Delta_{12} t} + \text{H.c.}\right]  \nonumber \\
	& + \frac{\eta_3}{2} (m^\dagger  + m ) { +  \frac{\varepsilon_1}{2} \cos(\omega_p t)(\ket{+}\bra{+} - \ket{-}\bra{-}). }\nonumber
\end{align}
By tuning the external drives such that $\eta_1 = \Delta_{12}$, we retain only the resonant terms in the above Hamiltonian.  Furthermore, assuming that the first drive is much stronger than the second one, i.e., $\eta_1 \gg \eta_2$, we neglect rapidly oscillating components and the Hamiltonian can be approximated as the following time-independent form:
\begin{align}\label{eq:H4}
	H_4 
	=&\ \Delta_m  m^\dagger m + \frac{\eta_2}{2}\frac{\sigma_z}{2} + \frac{G}{2} ( m + m^\dagger)\sigma_x \\
	+& \frac{\eta_3}{2} (m^\dagger  + m ) {+  \frac{\varepsilon_1}{2} \cos(\omega_p t)\sigma_x + \frac{\varepsilon_2}{4} \cos(\omega_p t)\sigma_z}.\nonumber
\end{align}
Applying a unitary displacement transformation $D = \exp[-\eta_3/(2\Delta_m) (m^\dagger - m) ]$ {and setting $\varepsilon_1 = \varepsilon_2/2 = 2\varepsilon$}, we can recast $H_4$ into a form closely resembling the quantum Rabi Hamiltonian:
\begin{align}\label{eq:H5}
	H_5&= \frac{\delta_z}{2}\sigma_z + \frac{\delta_x}{2} \sigma_x + \Delta_m m^\dagger m + \frac{G}{2} (m^\dagger + m ) \sigma_x \\
	&\ { +\,  \varepsilon \cos(\omega_p t)(\sigma_x+ \sigma_z)}\nonumber,
\end{align}
which includes a qubit tunneling term. Here $\delta_z = \eta_2/2$ is the longitudinal energy component associated with $\sigma_z$, and $\delta_x = -\eta_3 G /\Delta_m$ represents the effective transverse energy component associated with $\sigma_x$.

Since $\sigma_z$ and $\sigma_x$ do not commute, to diagonalize the qubit component, we introduce the dressed eigenstates $\ket{-} = \cos(\theta/2) \ket{\downarrow} - \sin(\theta/2) \ket{\uparrow}$ and $\ket{+} = \sin(\theta/2) \ket{\downarrow} + \cos(\theta/2) \ket{\uparrow}$, where $\theta = \arctan(\delta_z / \delta_x)$. The energy splitting between the dressed states is given by $\nu = \sqrt{\delta_z^2 + \delta_x^2}$. In the dressed-state basis, the Hamiltonian is written as
\begin{align}\label{eq:hxz}
	H_6 &= \frac\nu2\tilde{\sigma}_z + \Delta_m m^\dagger m + g_z (m^\dagger + m )\tilde{\sigma}_z 
															+ g_x (m^\dagger + m )\tilde{\sigma}_x  \nonumber\\
	    &\  { +\,  \varepsilon_p \cos(\omega_p t)\tilde{\sigma}_x}, 
\end{align}
where $\tilde{\sigma}_z = \ket{\uparrow}\bra{\uparrow} - \ket{\downarrow}\bra{\downarrow}$ and $\tilde{\sigma}_x = \ket{\uparrow}\bra{\downarrow} + \ket{\downarrow}\bra{\uparrow}$ and {$\varepsilon_p = \sqrt{2}\varepsilon/2$}. Notably, both the transverse coupling $g_x = G/2 \sin\theta$ and longitudinal coupling $g_z = G/2 \cos\theta$ are present in the magnon-qubit interaction. These couplings play a critical role in enabling a nonlinear two-magnon interaction with the qubit, which is essential for generating a magnonic cat state. 
Considering the above constraints, we take the following experimentally feasible parameters \cite{tabuchi2015coherent,lachance2017resolving,lachance2020entanglement,xu2023quantum}: 
{$\eta_1 = 2\pi\times 2.5$}~GHz, {$\eta_2 = 2\pi\times 50$}~MHz,  $g_x = g_z = \sqrt{2}G/4$  ($\theta = \frac{\pi}{4} $), $\omega_1 = \omega_3 = 2\pi \times 8.16734 $~GHz, {$\eta_3 = 2\pi \times 0.75 $}~MHz, and {$\omega_2 = 2\pi\times 5.66734 $}~GHz.

\section{Generation of magnon cat states \label{sec:3}}

In this section, we show how magnonic cat states can be generated in our system. The key element is to engineer a strong two-magnon nonlinear interaction, through which the qubit and the magnon mode exchange energy via magnon pairs.

To obtain the effective two-magnon interaction, we perform a unitary transformation $U = \exp[-i(m^\dagger m + \tilde{\sigma}_z)\frac{\omega_p}{2} t]$ on the Hamiltonian in Eq.~\eqref{eq:hxz}, which leads to the Hamiltonian
\begin{align}
	H_7 &= { \delta_m m^\dag m} + g_z (m e^{-i \frac{\nu}{2} t} + m^\dagger e^{i \frac{\nu}{2} t}) \tilde{\sigma}_z \\
	   &+ g_x (m \tilde{\sigma}_+ e^{i \frac{\nu}{2} t} {+} m^\dagger \tilde{\sigma}_+ e^{i \frac{3\nu}{2} t} {+} m \tilde{\sigma}_-  e^{-i \frac{3\nu}{2} t} {+} m^\dagger\tilde{\sigma}_- e^{-i \frac{\nu}{2} t}	)\nonumber\\
	   &+ \varepsilon_p (\tilde{\sigma}_+  + \tilde{\sigma}_- )\nonumber,
\end{align}
where $\delta_m = \Delta_m - \frac{\nu}{2}$ and we have assumed $ \omega_p = \nu$. In the regime $\nu \gg g_x, g_z, \varepsilon_p$, we employ the standard time-averaging method for highly detuned quantum systems to obtain the effective Hamiltonian \cite{james2007effective}. Under this condition, the resulting effective interaction can be expressed as
\begin{align}\label{eq:heff}
	H_{\text{eff}} &= \frac{8g_x^2}{3\nu} \left( \ket{\uparrow}\bra{\uparrow} + 2m^\dagger m \ket{\uparrow}\bra{\uparrow} \right)\\
	& - g_{\text{eff}} \left( m^2 \tilde{\sigma}_+ + m^{\dagger 2} \tilde{\sigma}_- \right) + \varepsilon_p\left( \tilde{\sigma}_+ + \tilde{\sigma}_- \right),\nonumber
\end{align}
where we have defined the effective coupling strength $g_\text{eff}=4g_x g_z/\nu$. The above Hamiltonian contains a nonlinear two-magnon interaction with the qubit. Unlike the conventional Jaynes-Cummings model, where single excitations are exchanged, this interaction allows the qubit and the magnon mode to exchange energy in the form of magnon pairs, providing a way to generate magnonic cat and squeezed states through the two-magnon process. We use achievable parameters {$\nu = \sqrt{2}/2\eta_2= 2\pi \times 35.4$}~MHz and the effective two-magnon coupling $g_\text{eff} = 2\pi\times 1.41 $~MHz, which we show later can efficiently generate the magnonic cat states. 

To prepare the magnonic cat state based on the effective Hamiltonian $H_{\text{eff}}$ in Eq.~\eqref{eq:heff}, we include the dissipation and decherence of the system. The time evolution of the system is described by the following master equation:
\begin{align}\label{eq:me}
\frac{d\rho}{dt} = -i[H_{\text{eff}}, \rho] + \frac{\gamma}{2} \mathcal{L}[\tilde{\sigma}_-]\rho
 + \frac{\gamma_\phi}{4} \mathcal{L}[\tilde{\sigma}_z]\rho + \frac{\kappa}{2} \mathcal{L}[m]\rho,
\end{align}
where $\rho$ is the density operator of the system, $\gamma$ and $\gamma_\phi$ denote the energy relaxation and pure dephasing rates of the qubit, respectively, and $\kappa$ is the dissipation rate of the magnon mode. Here, $\mathcal{L}[o]\rho = 2o\rho o^\dagger - o^\dagger o \rho - \rho o^\dagger o$ is the standard Lindblad superoperator for a given operator $o$, where $o = \{\tilde{\sigma}_-, \tilde{\sigma}_z, m\}$. 
To express more clearly our dissipation mechanism for preparing magnonic cat states, we first neglect the dephasing (dissipation) of the qubit (magnon), i.e., $\gamma_\phi=\kappa=0$, of which their effects will be studied later.

Equation~\eqref{eq:me} describes a two-magnon driving-dissipation process that governs the system's dynamics. The driving field resonantly drives the qubit to its excited state. Upon reaching the excited state, the nonlinear interaction $\sim g_{\text{eff}}(m^2 \tilde{\sigma}_+ + m^{\dagger 2} \tilde{\sigma}_-)$ tells that the qubit jumps to its ground state by emitting a pair of magnons to the magnon mode (corresponding to the driving of the magnon mode). Conversely, the qubit jumps to its excited state by absorbing a pair of magnons (corresponding to the dissipation of the magnon mode), and the qubit subsequently dissipates the energy to the environment via spontaneous emission. The system evolves to a steady state when the above two processes achieve dynamic equilibrium, which leads to a magnon cat state \cite{gerry1993generation}. Specifically, due to the driving-dissipation competition, the system evolves toward a steady state, satisfying $d\rho_{ss}/dt = 0$, where $\rho_{ss} = |\Psi_{ss}\rangle \langle \Psi_{ss}|$, $|\Psi_{ss}\rangle = |\phi_{ss}\rangle \otimes |\psi_{ss}\rangle$, and $|\phi_{ss}\rangle$ and $|\psi_{ss}\rangle$ denote the steady states of the qubit and the magnon mode, respectively. The steady state of the qubit $\ket{\phi_{ss}}$ is in the ground state $\ket{\downarrow}$ and the magnon steady state $|\psi_{ss}\rangle$ satisfies the equation $H_{\text{eff}}\ket{\psi_{ss}} \otimes \ket{\downarrow} = 0$. Solving this, the steady state of the magnon mode is given by (Appendix \ref{appendix})
\begin{align}\label{eq:catstate}
	\ket{\psi_{ss}} = c_+\ket{\alpha} + c_-\ket{-\alpha},
\end{align}
which is the superposition of two coherent states, i.e., a cat state.
Here, $\alpha = \sqrt{\varepsilon_p / g_\text{eff}}$ is the complex amplitude, and $c_\pm $ are two constants. 
Since the qubit's steady state is the ground state, this approach is robust against qubit dephasing.

\begin{figure}[b]
	\centering
	\includegraphics[width=0.96\linewidth]{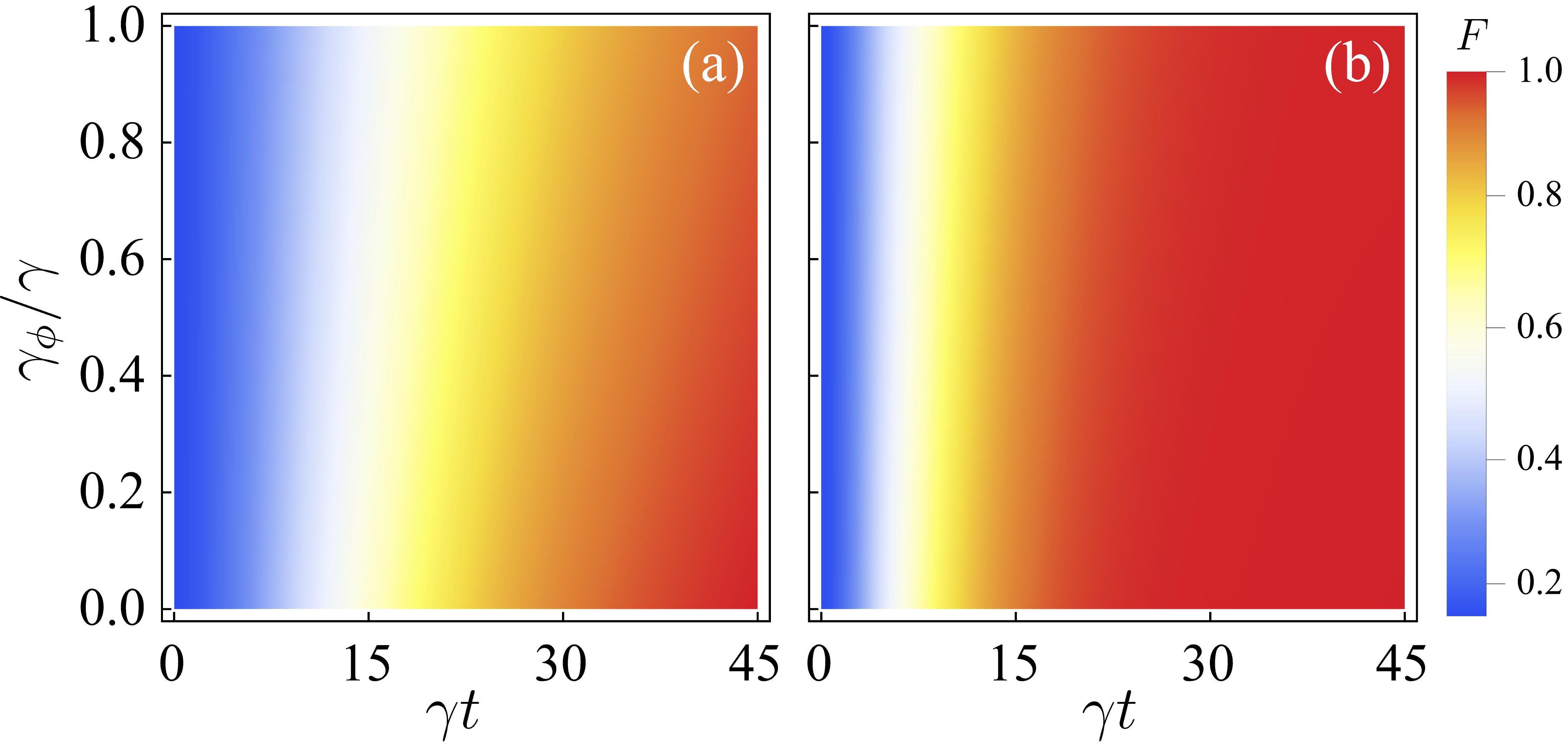}
	\caption{
		The fidelity $ F $ as a function of the dimensionless time $ \gamma t $ and qubit dephasing rates $ \gamma_\phi/\gamma $. (a) The even cat state, with the initial state of the system being $ \ket{0} \otimes \ket{\downarrow} $; (b) The odd cat state, with the initial state of the system as $ \ket{1} \otimes \ket{\downarrow} $. We take $\gamma = 2\pi \times 16 \, \text{MHz}$, {$\varepsilon_p = 2\pi \times 3.53 $}~MHz, and $\kappa=0$. See text for the other parameters, which lead to a cat state with the amplitude $\alpha = 1.58$ in the steady state.
	}
	\label{fig:fidelity}
\end{figure}

\begin{figure*}[t]
	\centering
	\includegraphics[width=0.98\linewidth]{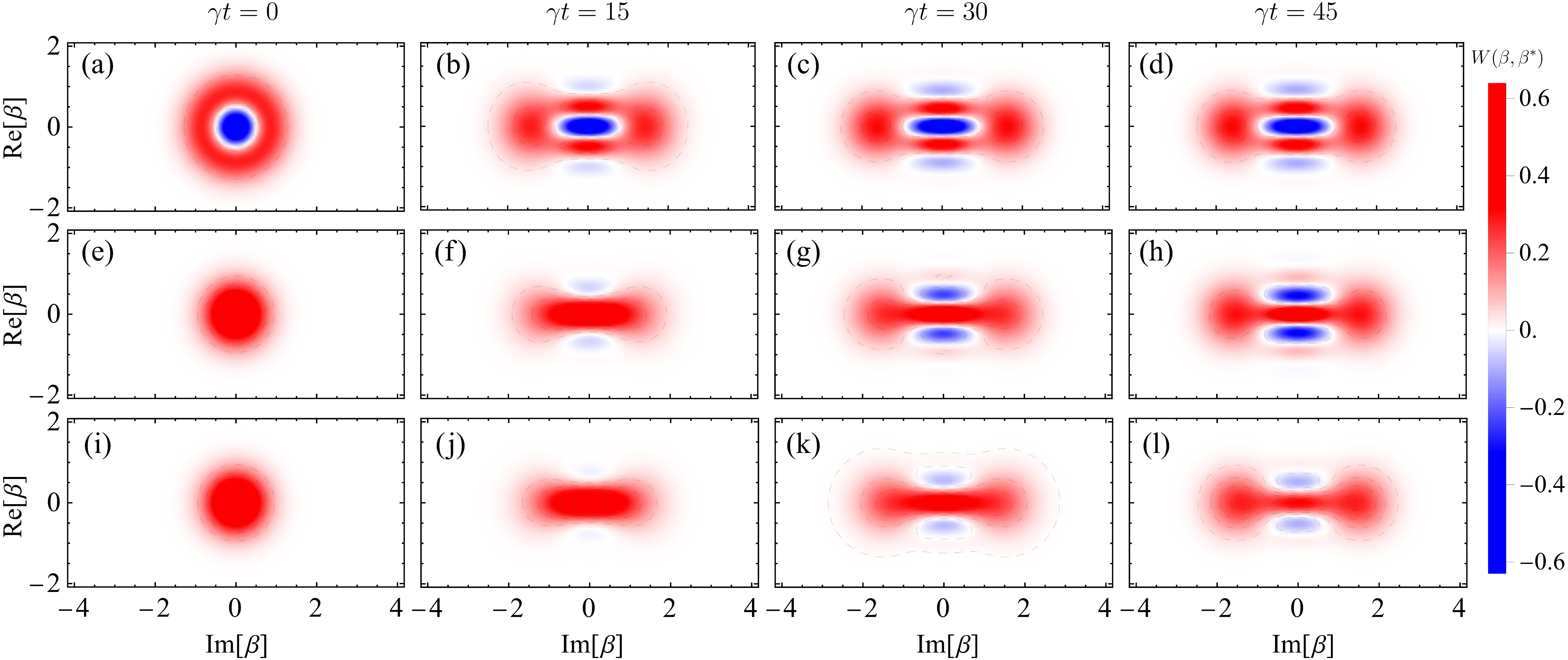}
	\caption{Time evolution of the Wigner function for an odd cat state (a)-(d) and an even cat state (e)-(h) with $\kappa=0$; and for an even cat state (i)-(l) with $\kappa/2\pi = 0.5$~MHz. Initially, there are no interference fringes, but they gradually develop over time, leading to an increase in fidelity (cf. Fig.~\ref{fig:fidelity}). The parameters are the same as those in Fig.~\ref{fig:fidelity} and we take $\gamma_\phi = \gamma$.}
	\label{fig:ideal_cat}
\end{figure*}

Since the dynamics governed by the master equation conserves the magnon-number parity, defined by the operator $P = e^{i\pi m^\dagger m}$, the parity of the steady state of the magnon mode depends on the parity of its initial state. If the magnon mode is initially prepared in an even-parity state (e.g., the vacuum state $|0\rangle$), then $c_+ = c_-$, and the magnon steady state corresponds to an even cat state
\begin{align}
	\ket{\psi_{ss}^e} = \mathcal{N}_+(\ket{\alpha} + \ket{-\alpha}).
\end{align}
For an initial odd-parity state (e.g., the single-magnon state $|1\rangle$), we have $c_+ = - c_-$ and the magnon steady state is in an odd cat state
\begin{align}
	\ket{\psi_{ss}^o} = \mathcal{N}_-(\ket{\alpha} - \ket{-\alpha}).
\end{align}
Furthermore, the amplitude $\alpha$ can be precisely controlled by tuning the driving microwave field.

As analyzed above, the two-magnon driving-dissipation mechanism for magnonic cat states requires the dissipation of the qubit. In what follows, we further study the impact of other decoherence channels, i.e., the qubit dephasing and magnon dissipation, on the target state Eq.~\eqref{eq:catstate}. We first study the effect of the qubit dephasing by numerically solving Eq.~\eqref{eq:me} with the effective Hamiltonian Eq.~\eqref{eq:heff} and a nonzero qubit dephasing rate $\gamma_\phi$ but a zero magnon dissipation rate $\kappa=0$.
Here we define the fidelity $F = \text{Tr}[\rho_\text{tar} \rho_m]$ to quantify the overlap between the evolved state $\rho_m$ and the target state $\rho_\text{tar} = \ket{\psi_{ss}}\bra{\psi_{ss}}$ \cite{ourjoumtsev2007generation,branczyk2008teleportation,sychev2017enlargement}, where $\rho_m = \text{Tr}_\text{qubit}[\rho]$ is the reduced density matrix of the magnon mode obtained by tracing out the qubit degrees of freedom. 
Figure \ref{fig:fidelity} shows the time evolution of the fidelity $ F $ versus the dimensionless time $ \gamma t $ and the dephasing rate $ \gamma_\phi $ for an even (odd) cat state in Fig.~\ref{fig:fidelity}(a) [Fig.~\ref{fig:fidelity}(b)], corresponding to the initial magnon state being a vacuum (single-magnon) state. Clearly, for both cases the dephasing rate does not degrade the final fidelity, but only slightly increases the time to reach a unity fidelity.
The negligible impact of the qubit dephasing is due to the fact that, as analyzed before, the qubit eventually relaxes to its ground state in the steady state. This, in turn, reflects the robustness of our scheme against the qubit dephasing. In the following discussion, we set $ \gamma_\phi = \gamma $ unless otherwise specified.

\begin{figure}[t]
	\centering
	\includegraphics[width=0.96\linewidth]{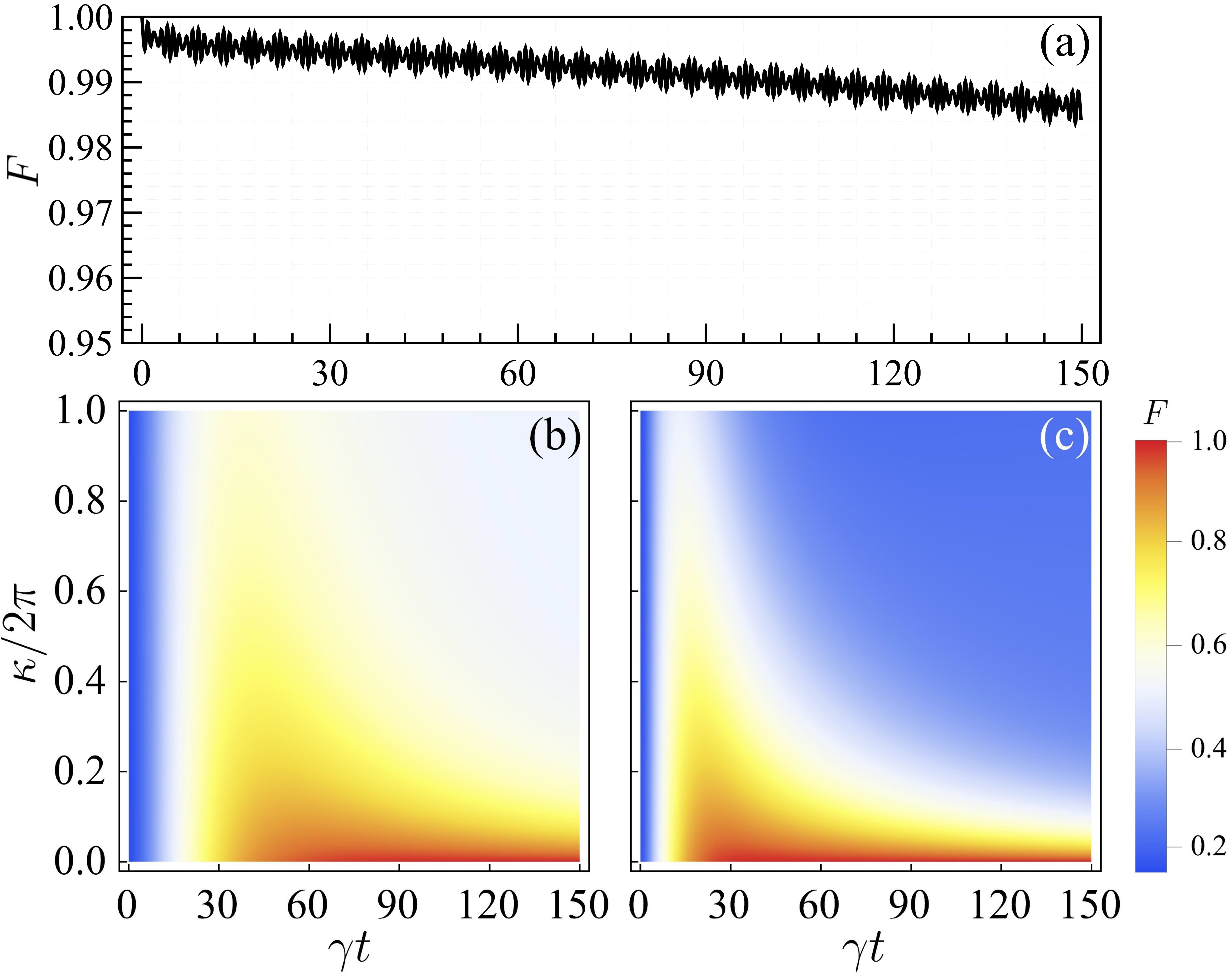}
	\caption{(a) The fidelity $F$ between two evolved states governed by the effective Hamiltonian Eq.~\eqref{eq:heff} and the full Hamiltonian Eq.~\eqref{eq:full}  by solving Eq.~\eqref{eq:me} with the initial state $\ket{0} \otimes \ket{\downarrow}$, $ \gamma = 2\pi \times 16~\text{MHz} $, and $ \gamma_\phi = \kappa = 0 $, as a function of $ \gamma t $.
		(b)-(c) The fidelity $ F $ versus $ \gamma t $ and magnon dissipation rate $ \kappa $, obtained by numerically solving Eq.~\eqref{eq:me} with the magnon initial state (b) $ \ket{0}$; and (c) $ \ket{1} $. The parameters are the same as those in Fig.~\ref{fig:fidelity}.
	}
	\label{fig:fidelity3D1}
\end{figure}

Figure \ref{fig:ideal_cat} shows the evolution of the Wigner function of the magnon mode by including the magnon dissipation. The Wigner function is defined as
\begin{equation}
	W(\beta) = \frac{2}{\pi} \text{Tr}[e^{i\pi m^\dag m}\tilde{\rho}],
\end{equation}
where $\tilde{\rho}_m = D(\beta)\rho_m D^{\dagger}(\beta) $, and $D(\beta) = e^{\beta m^{\dagger} - \beta^* m}$ is the displacement operator. 
Specifically, Fig.~\ref{fig:ideal_cat} shows the evolution of the Wigner function for two situations: $i$) without magnon dissipation [panels (a)-(d) for the odd cat state; (e)-(h) for the even cat state], and $ii$) with magnon dissipation $\kappa/2\pi = 0.5$~MHz [panels (i)-(l) for the even cat state]. Clearly, for both cases of the magnon initial state ($\ket{0}$ or $\ket{1}$), the magnon mode evolves into a cat state. For the case of $\kappa=0$, the final cat state is exactly the target state. Panels (i)-(l) of Fig.~\ref{fig:ideal_cat} indicate that the magnon dissipation can significantly reduce the fidelity to the desired state. For a low dissipation rate $\kappa/2\pi \approx 0.5$~MHz of a YIG sphere used in the experiment \cite{Shen2023arxiv}, we obtain a maximum fidelity {$F \approx 0.7$ at $\gamma t = 41$}. This is because the energy dissipation of the magnon mode breaks parity conservation during the quantum state preparation, leading to the loss of coherence. Nonetheless, the impact of the magnon dissipation can be mitigated by increasing the nonlinear coupling $g_{\rm eff}$.
In Fig.~\ref{fig:fidelity3D1}(a), we plot the fidelity $ F $ between the two evolved states governed by the derived effective Hamiltonian Eq. \eqref{eq:heff} and the full Hamiltonian Eq. \eqref{eq:full}  by solving the master equation with the initial state $\ket{0} \otimes \ket{\downarrow}$, $ \gamma = 2\pi \times 16~\text{MHz} $, and $ \gamma_\phi = \kappa = 0 $. It shows that the effective Hamiltonian Eq. \eqref{eq:heff} is a good approximation of the full Hamiltonian (with $F>92\%$ for $\gamma t \leq 45$).
In Fig.~\ref{fig:fidelity3D1}(b)-(c), we also plot the fidelity $ F $ between the evolved state $\rho_m$ (under the Hamiltonian Eq. \eqref{eq:heff}) and the target state $\rho_\text{tar}$ as a function of $ \gamma t $ and magnon dissipation rate $ \kappa$. It shows that for a very small $ \kappa$, the fidelity is close to unity in the long-time limit, which implies that the aforesaid two-magnon driving-dissipation process is dominant, enabling the generation of long-lived target states.  As $ \kappa$ increases, the fidelity, after reaching a maximum, decreases over time, reflecting the loss of coherence due to the considerable magnon dissipation.
{In addition, we study the effect of the experimental imperfections, e.g., the mismatch of driving frequencies and strengths, on the fidelity in Appendix~\ref{appendixB}. Further, we show in Appendix~\ref{appendixC} that there is a trade-off between the fidelity and the size of the cat state when optimizing the effective coupling $g_\text{eff}$.}



\begin{figure}[t]
	\centering
	\includegraphics[width=0.98\linewidth]{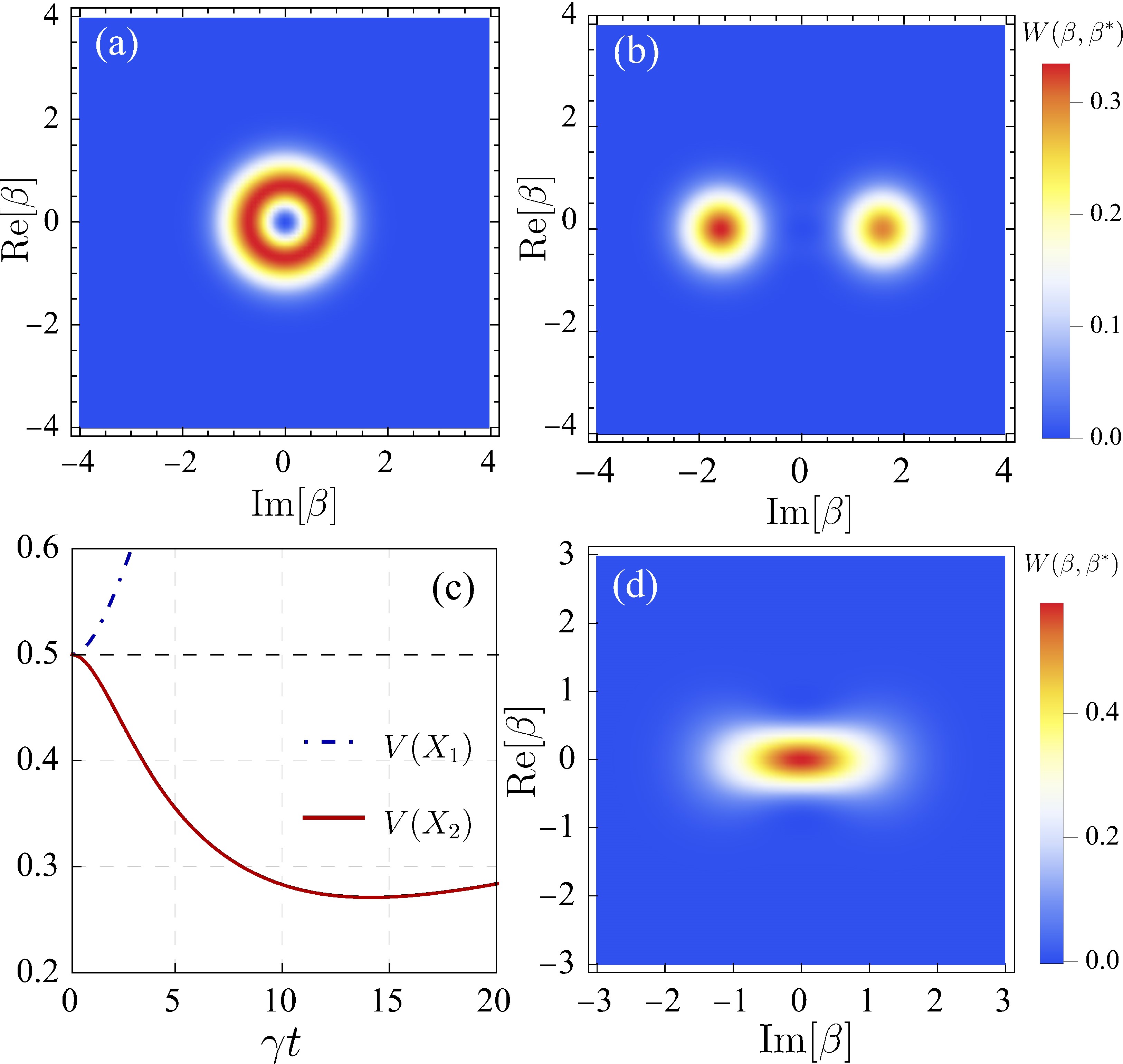}
	\caption{Wigner function of (a) the initial state $ \ket{\psi(0)} = \frac{1}{\sqrt{2}} (\ket{0} + i\ket{1}) $ and of (b) the steady state of the magnon mode by numerically solving Eq.~\eqref{eq:me} in the absence of magnon dissipation. 
		(c)  Variance of two quadratures $ X_1 $ and $ X_2 $ of the magnon mode versus $\gamma t$, and (d) Wigner function corresponding to the minimum variance of $X_2$ in (c). The dashed horizontal line denotes the variance of vacuum fluctuation. For (c)-(d), the magnon mode is initially in vacuum state $\ket{0}$ and $\kappa/2\pi = 0.5$~MHz. The other parameters are identical to those in Fig.~\ref{fig:fidelity}. 
	}
	\label{fig:mixSqeezed}
\end{figure}

The dynamics without magnon dissipation preserves the parity of the magnon excitation number, and thus the parity of the initial magnon state determines that of the steady state. Specifically, if the initial state is either the vacuum $ \ket{0} $ or the single-magnon state $ \ket{1} $, the steady state corresponds to the even cat state $ \ket{\psi^e} $ or the odd cat state $ \ket{\psi^o} $, respectively. For an initial superposition state, e.g., $ \ket{\psi(0)} = \cos(\theta/2)\ket{0} + \sin(\theta/2)e^{i\phi}\ket{1} $~\cite{xu2023quantum}, where $ 0 \leq \theta \leq \pi $ and $ 0 \leq \phi < 2\pi $, the steady state becomes a weighted mixture of the even and odd cat states. 
The proportion of the components is determined by the initial state (more details see Appendix \ref{appendix}). Specifically, for $ \theta = 0 $, the initial state corresponds to the vacuum $ \ket{0} $, and the steady state is the even cat state $ \ket{\psi^e}\bra{\psi^e} $; for $ \theta = \pi $, the initial state corresponds to the single-magnon state $ \ket{1} $, and the steady state is the odd cat state $ \ket{\psi^o}\bra{\psi^o} $. For $ 0 < \theta < \pi/2 $, the probability of the even cat state is greater than that of the odd cat state, while for $ \pi/2 < \theta < \pi $, the opposite. Particularly, for $ \phi = \pi/2, 3\pi/2 $, the probabilities of $ \ket{\psi^e} \bra{\psi^o} $ and $ \ket{\psi^o} \bra{\psi^e} $ are equal, leading to the cancellation of interference fringes (Appendix \ref{appendix}). Further, when $ \theta = \pi/2 $, the probabilities of the even and odd cat states are equal, leading to an equal mixture of both, i.e., $ \ket{\psi^e}\bra{\psi^e} + \ket{\psi^o}\bra{\psi^o} $. Specifically, Fig.~\ref{fig:mixSqeezed}(a) illustrates the magnon initial state $\ket{\psi(0)} =\frac{1}{\sqrt{2}} (\ket{0} + i\ket{1}) $, which leads to the mixed steady state $\ket{\alpha}\langle\alpha| + \ket{-\alpha}\langle -\alpha|$, as shown in Fig.~\ref{fig:mixSqeezed}(b).
It is noteworthy that at short times, the two-magnon driving process is dominant, which can induce squeezing of the magnon mode, as confirmed by Fig.~\ref{fig:mixSqeezed}(c). It is clear that one of the two quadratures $X_1$, $X_2$ of the magnon mode is squeezed, where $X_1 = (m + m^\dag)/\sqrt{2}$ and $X_2 = i( m^\dag - m)/\sqrt{2}$,
as embodied by its variance $V(X_2)$ below that of the vacuum fluctuation.  
Figure \ref{fig:mixSqeezed}(d) shows the Wigner function of the magnon mode at the time giving the minimum variance of $X_2$.
This reflects the rich dynamics of our two-magnon process in preparing nonclassical magnon states either dynamically or in the steady state.

\section{Conclusion \label{sec:4}}

We have proposed an efficient approach for preparing a long-lived cat state or a dynamical squeezed state of the magnon mode, based on a two-magnon driving and dissipation mechanism. The protocol relies on the nonlinear two-magnon interaction induced by appropriately selected multiple drives applied to the qubit and magnon mode. The qubit dissipation is a necessary element of our mechanism. Although the qubit dephasing does not affect the final fidelity to the desired cat state, the magnon dissipation can considerably reduce the fidelity in the long-time limit. {
Compared to previous methods \cite{sun2021prl,he2023pramc,he2024pracats}, which generate transient cat states, our protocol produces steady cat states. Although a different scheme was offered for preparing steady cat states \cite{liu2024magnoncat}, it requires parametric coupling between two microwave cavities, which remains experimentally challenging.}
Our method provides a robust and flexible platform for generating nonclassical magnonic states and may find applications in quantum information processing and quantum sensing, including dark matter detection using ferromagnetic axion haloscopes~\cite{flower2019broadening,crescini2020axion}.

\begin{acknowledgments}
	
We thank G. S. Agarwal for useful discussion. This work was supported by Zhejiang Provincial Natural Science Foundation of China (Grant No. LR25A050001), National Natural Science Foundation of China (Grant No. 12474365, 12174140, 92265202) and National Key Research and Development Program of China (Grant No. 2022YFA1405200, 2024YFA1408900).
	
\end{acknowledgments}


\appendix 

\textcolor{black}{
\section{Derivation of the steady-state solution \label{appendix}}}

Here, we provide a detailed derivation of the steady state solution of Eq.~\eqref{eq:me} for $\gamma_\phi=\kappa=0$, i.e., including only the qubit dissipation. The master equation then becomes
\begin{align}
	\frac{d\rho}{dt} = -i[H_{\text{eff}}, \rho] + \frac{\gamma}{2} \mathcal{L}[\tilde{\sigma}_-]\rho.
\end{align}
For the steady state $\rho_{ss}$, we have
\begin{align}
	-i[H_{\text{eff}}, \rho_{ss}] + \frac{\gamma}{2} \mathcal{L}[\tilde{\sigma}_-]\rho_{ss} = 0.
\end{align}
Assuming the steady state is a pure state $\rho_{ss} = |\Psi_{ss}\rangle \langle \Psi_{ss}|$, here we express $\ket{\Psi_{ss}}$ as a separable state, i.e., $\ket{\Psi_{ss}} = \ket{\phi_{ss}} \otimes \ket{\psi_{ss}}$, with $|\phi_{ss}\rangle$ and $|\psi_{ss}\rangle$ being the steady states of the qubit and magnon mode, respectively. Substituting this into the above steady-state equation, we obtain
\begin{align}
	-i[H_{\text{eff}}^{\text{nh}}, \rho_{ss}] + \gamma \tilde{\sigma}_- \rho \tilde{\sigma}_+ = 0,
\end{align}
where the non-Hermitian Hamiltonian $H_{\text{eff}}^{\text{nh}} = H_{\text{eff}} - i \frac{\gamma}{2}\tilde{\sigma}_+\tilde{\sigma}_- $. 
%
%
By analyzing the equation, we find that the solution satisfies the conditions $ \tilde{\sigma}_- \rho_{ss} \tilde{\sigma}_+ = 0$ and $[H_{\text{eff}}^{\text{nh}}, \rho_{ss}] = 0$. Solving the former, we get $\ket{\phi_{ss}} = \ket{\downarrow}$.
It is easy to obtain that the solution must satisfy the following:
\begin{align}
	m^2 \ket{\psi_{ss}}\bra{\psi_{ss}} &= z\ket{\psi_{ss}}\bra{\psi_{ss}},\\
	\ket{\psi_{ss}}\bra{\psi_{ss}}m^{\dag2} &= \ket{\psi_{ss}}\bra{\psi _{ss}}z^*,
\end{align}
with $z = \Omega/g_\text{eff}$.  Clearly, the possible states of the ensemble $\ket{\psi_{ss}}$ must satisfy
\begin{align}
	m^2 \ket{\psi_{ss}} = z\ket{\psi_{ss}}, \bra{\psi_{ss}}m^{\dag2} = \bra{\psi_{ss}}z^*,
\end{align}
which lead to the following general form:
\begin{align}\label{aaa8}
	\ket{\psi_{ss}} = c_+ \ket{\alpha} + c_- \ket{-\alpha},
\end{align}
where $\alpha^2 =z = \Omega/g_\text{eff}$.

Next, we expand $\ket{\psi_{ss}}$ in the Fock-state basis as $\ket{\psi_{ss}} = \sum_n c_n \ket{n}$ ($n \geq 0$), with $c_n$ being the probability amplitude. Substituting this into Eq.~\eqref{aaa8} yields the recursion relation
\begin{align}\label{eq:recursion}
	c_{n+2} = \frac{\alpha^2}{\sqrt{(n+1)(n+2)}} c_n,
\end{align}
which indicates that when the magnon is initially in a Fock state $\ket{k}$ with an even $k=2n$, 
then $c_+ = c_-$, and the steady-state $\ket{\psi_{ss}}$ can be expressed as
\begin{align}
	\ket{\psi^e} &= \sqrt{\frac{2}{e^{|\alpha|^2} + e^{-|\alpha|^2}}} \sum_{n=\text{even}} \frac{\alpha^n}{\sqrt{n!}} \ket{n}\nonumber\\
	&=\mathcal{N}_+ (\ket{\alpha}+\ket{-\alpha}),
\end{align}
which is an even cat state. Similarly, for the magnon initial state $\ket{k}$ with an odd $k=2n+1$, 
then $c_+ = - c_-$, and the steady state is in the form
\begin{align}
	\ket{\psi^o} &= \sqrt{\frac{2}{ e^{|\alpha|^2} - e^{-|\alpha|^2} }} \sum_{n=\text{odd}} \frac{\alpha^n}{\sqrt{n!}} \ket{n}\nonumber\\
	&=\mathcal{N}_- (\ket{\alpha}-\ket{-\alpha}),
\end{align} 
which is an odd cat state.
The normalization constants are given by $\mathcal{N}_\pm =  [2(1\pm e^{-2|\alpha|^2})]^{-1/2}$. 

For the magnon initial state being a superposition state
\begin{align}\label{eq:init}
	\ket{\psi(0)} = a \ket{2n} + b\ket{2n+1},
\end{align}
where $|a|^2+|b|^2 = 1$. The steady state is then the superposition of an even and an odd cat state, i.e.,
\begin{align}
	\ket{\psi_{ss}} = a \ket{\psi^e} + b \ket{\psi^o}.
\end{align}
The corresponding density operator $\rho_{ss}$ is then
\begin{align}\label{eq:ss}
	\rho_{ss} 
	&= |a|^2 \ket{\psi^e} \bra{\psi^e} + a^* b  \ket{\psi^e} \bra{\psi^o}\\
	&\, + ab^* \ket{\psi^o} \bra{\psi^e} +  |b|^2 \ket{\psi^o} \bra{\psi^o}\nonumber.
\end{align}
To be specific, we consider the following magnon initial state, which has been experimentally achieved~\cite{xu2023quantum}
\begin{align}\label{eq:super}
	\ket{\psi(0)} = \cos\left(\frac{\theta}{2}\right) \ket{0} + \sin\left(\frac{\theta}{2}\right)e^{i\phi}\ket{1},
\end{align}
where $0 \le \theta \le \pi, 0 \le \phi < 2\pi$. This corresponds to $a = \cos(\theta/2)$, $b = \sin(\theta/2)e^{i\phi}$, and $n=0$ in Eq.~\eqref{eq:init}. 
Clearly, for $ \theta = 0$ ($\pi$), the initial state is the vacuum (single-magnon) state, corresponding to the magnon steady state being an even (odd) cat state. For other cases, the steady state takes a general form as in Eq.~\eqref{eq:ss}.  For a special case of $ \theta = \pi/2$, the initial state is a superposition state of $\ket{0}$ and $\ket{1}$ with equal probability, and when further $\phi = \pi/2$ or $3\pi/2$, $a^* b=-ab^*$ and thereby the two interference terms in Eq.~\eqref{eq:ss} cancel out, leading to an equal mixture of both the even and odd cat states, i.e., $ \ket{\psi^e}\bra{\psi^e} + \ket{\psi^o}\bra{\psi^o}$, whose Wigner function has no interference fringes, as shown in Fig.~\ref{fig:mixSqeezed}(b).

{\section{Effect of the mismatch of driving frequencies and strengths} \label{appendixB}}

\begin{figure}[tp] 
	\centering
	\includegraphics[width=0.86\linewidth]{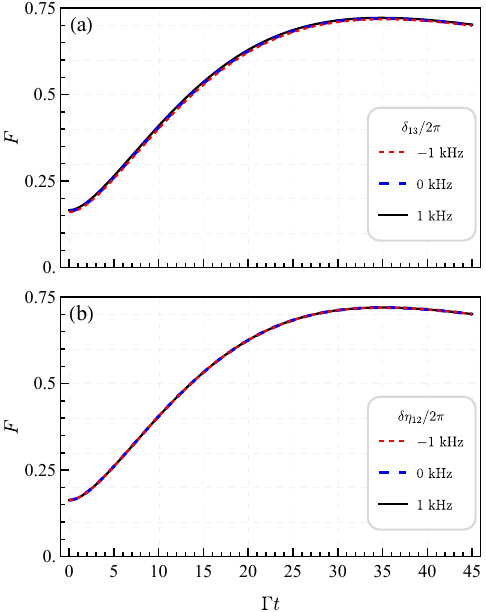}
	\caption{{(a) The fidelity versus $\Gamma t$ for three values of $\delta_{13}$: $\delta_{13}/2\pi = -1\ \text{kHz}$ (red dashed), 0 (black solid), and $1\ \text{kHz}$ (blue dashed). (b) The fidelity versus $\Gamma t$ for three values of $\delta\eta_{{12}} = \eta_1 - (\omega_1 - \omega_2)$. The line styles and the corresponding detunings are the same as in (a). The other parameters are identical to those in Fig.~\ref{fig:fidelity}}.} 
	\label{fig:appB}
\end{figure}

{In deriving the effective Hamiltonian Eq.~\eqref{eq:H4}, we ideally assumed $\omega_1 = \omega_3$ and $\eta_1 = \omega_1 - \omega_2$. However, in real experiments, there might be mismatch between the driving frequencies and strengths. In what follows, we study the impact of such mismatch on the fidelity of our proposed cat state.}

{We first study the effect of the frequency mismatch $\delta_{{13}} = \omega_1 - \omega_3 \neq 0$, which introduces a residual time-dependent drive on the magnon mode. This effect can be absorbed via a time-dependent displacement transformation, leading to a renormalized transverse qubit coupling $\delta_x = -\eta_3 G/(\Delta_m - \delta_{{13}})$. The resulting Hamiltonian retains the same form as Eq.~\eqref{eq:H5}, with slightly modified parameters. Figure \ref{fig:appB}(a) shows the effect of such frequency mismatch up to 1 kHz. Clearly, the fidelity changes only slightly, indicating that our protocol is  robust against the mismatch of driving frequencies.
	}

{We also study the effect of the deviation from the perfect drive-matching condition $\eta_1 = \omega_1 - \omega_2$, i.e., $\delta\eta_{12}= \eta_1 - (\omega_1 - \omega_2)  \neq 0$. Figure \ref{fig:appB}(b) shows the effect of such mismatch up to 1 kHz on the fidelity. It is clear that the fidelity almost remains the same, confirming the robustness of the protocol towards the mismatch of the driving conditions.
	}

{\section{Trade-off of the effective coupling strength $g_\text{eff}$} \label{appendixC}}

\begin{figure}[hb] 
	\centering
	\includegraphics[width=0.86\linewidth]{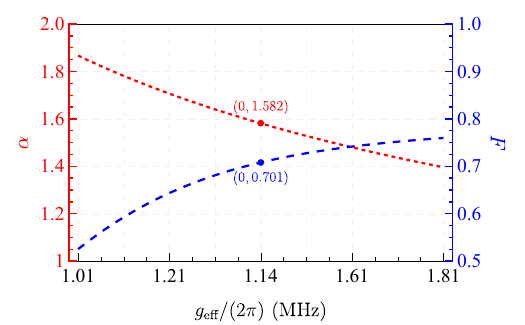}
	\caption{{Fidelity (blue) and coherent amplitude $|\alpha|$ (red) versus the effective coupling strength $g_\text{eff}$. The other parameters are identical to those in Fig.~\ref{fig:fidelity}. The data correspond to the steady state at time $\Gamma t = 45$.}}
	\label{fig:appC}
\end{figure}

{
The effective coupling strength $g_\text{eff}$ is a key parameter in our protocol and it is thus important to optimize it. We note that $g_\text{eff}$ is not necessarily the larger the better, because there is a trade-off between the fidelity and the size of the cat state. While a larger effective coupling (e.g., by increasing bare couplings or optimizing detunings) improves the fidelity of the cat state (see the blue curve in Fig.~\ref{fig:appC}), it simultaneously reduces the size of the cat state, since $|\alpha| \propto g_\text{eff}^{-1/2}$ (see the red curve in Fig.~\ref{fig:appC}). Therefore, to determine an optimal coupling strength, one should consider two factors, i.e., the fidelity and the size of the cat state.
}

\bibliographystyle{apsrev4-2} 
\bibliography{ref}

\end{document}